\documentclass[12pt]{iopart}
\usepackage{graphicx}

\begin{document}

\title{First-order transition features of the 3D bimodal
random-field Ising model}

\author{N G Fytas, A Malakis and K Eftaxias}

\address{Department of Physics, Section of Solid State
Physics, University of Athens, Panepistimiopolis, GR 15784
Zografos, Athens, Greece} \eads{\mailto{amalakis@phys.uoa.gr}}

\begin{abstract}
Two numerical strategies based on the Wang-Landau and Lee entropic
sampling schemes are implemented to investigate the first-order
transition features of the 3D bimodal ($\pm h$) random-field Ising
model at the strong disorder regime. We consider simple cubic
lattices with linear sizes in the range $L=4-32$ and simulate the
system for two values of the disorder strength: $h=2$ and
$h=2.25$. The nature of the transition is elucidated by applying
the Lee-Kosterlitz free-energy barrier method. Our results
indicate that, despite the strong first-order-like
characteristics, the transition remains continuous, in
disagreement with the early mean-field theory prediction of a
tricritical point at high values of the random-field.
\end{abstract}

 \vspace{2pc}

\maketitle

\section{Introduction}
\label{sec:1}

The random-field Ising model
(RFIM)~\cite{imry75,aharony76,young77,parisi78,grinstein82,
imbrie84,villain84,schwartz85,bray85,houghton85,young85,fisher86,binder86,ogielski86,bricmont87}
has been extensively studied both because of its interest as a
simple frustrated system and because of its relevance to
experiments~\cite{belanger98}. The Hamiltonian describing the
model is
\begin{equation}
\label{eq:1}
\mathcal{H}=-J\sum_{<i,j>}S_{i}S_{j}-h\sum_{i}h_{i}S_{i},
\end{equation}
where $S_{i}$ are Ising spins, $J>0$ is the nearest-neighbors
ferromagnetic interaction, $h$ is the disorder strength, also
called randomness of the system, and $h_{i}$ are independent
quenched random-fields (RF's) obtained here from a bimodal
distribution of the form
\begin{equation}
\label{eq:2} P(h_{i})=\frac{1}{2}[\delta(h_{i}-1)+\delta(h_{i}+1].
\end{equation}
Various RF probability distributions, such as the Gaussian, the
wide bimodal distribution (with a Gaussian width), and the above
bimodal distribution have been
considered~\cite{gofman93,cao93,falicov95,swift97,auriac97,hartmann99,sourlas99,machta00,hartmann01,middleton02,wu05,wu06,fytas07}.

As it is well known, the existence of an ordered ferromagnetic
phase for the RFIM, at low temperature and weak disorder, follows
from the seminal discussion of Imry and Ma~\cite{imry75}, when
$D>2$. This has provided us with a general qualitative agreement
on the sketch of the phase boundary separating the ordered
ferromagnetic (\textbf{F}) phase from the high-temperature
paramagnetic (\textbf{P}) phase.
\begin{figure}[ht]
\centerline{\includegraphics*[width=12 cm]{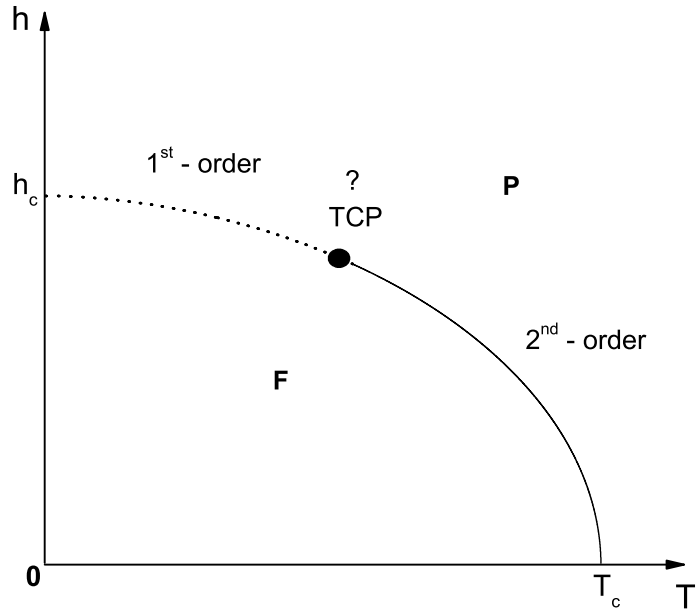}}
\caption{\label{fig:1}A sketch of the phase boundary of the 3D
bimodal RFIM, where $h_{c}$ is the critical disorder strength and
$T_{c}$ the critical temperature of the pure 3D Ising model. The
question-mark refers to the mean-field prediction of a tricritical
point (TCP), where the transition supposedly changes from
second-order at low-fields (solid line) to first-order at
high-fields (dotted line).}
\end{figure}
The phase boundary (see figure~\ref{fig:1}) separates the two
phases of the model and intersects the randomness axis at the
critical value of the disorder strength $h_{c}$. This value of
$h_{c}$ is known with good accuracy for both the Gaussian and the
bimodal RFIM to be $2.270(4)$~\cite{middleton02} and
2.21(1)~\cite{auriac97,hartmann99,fytas07}, respectively. A most
recent detailed numerical investigation of the phase boundary of
the 3D bimodal RFIM appears in reference~\cite{fytas07}.

However, the general behavior of phases and phase transitions in
systems with quenched randomness is still
controversial~\cite{harris74,berker84,dotsenko07}, and one such
lively example is the 3D RFIM, which, despite $30$ years of
theoretical and experimental study, is not yet well understood. In
particular, the nature of its phase transition remains unsettled,
although it is generally believed that the transition from the
ordered to the disordered phase is continuous governed by the
zero-temperature random
fixed-point~\cite{villain84,bray85,fisher86}. For the bimodal
RFIM, the mean-field prediction~\cite{aharony78} of a first-order
region separated from a second-order region by a TCP, remains
today an open controversy. This main issue has regained interest
after the recent observations~\cite{hernandez97,hernandez07} of
first-order-like features at the strong disorder regime. Nowadays,
this is the main conflict regarding the nature of the phase
transition of the 3D bimodal RFIM, although other controversies
and scenarios exist in the literature, concerning mainly the
intermediate regime of the phase diagram and a possible third
spin-glass phase~\cite{mezard92,brezin98,sinova01}.

Thus, the possibility of a first-order transition at the strong
disorder regime has been discussed in several papers and has been
supported over the years by numerical and theoretical findings.
The extreme sharpness of the transition reflected in some studies
in the estimated very small values of the order-parameter exponent
$\beta$~\cite{falicov95,middleton02} has also been reinforcing
such first-order scenarios. In particular first-order-like
features, such as the appearance of the characteristic double-peak
(dp) structure of the canonical energy probability density
function (PDF), have been recently reported for both the Gaussian
and the bimodal distributions of the 3D RFIM. Particularly, Wu and
Machta~\cite{wu06}, using the Wang-Landau (WL)
approach~\cite{wang01a,wang01b,schulz03}, reported such properties
for the Gaussian RFIM at a strong disorder strength value $h=2$
below their critical randomness ($h_{c}=2.282$). Moreover,
Hern\'{a}ndez and Diep~\cite{hernandez97} have emphasized that
they have found evidence for the existence of a TCP in the phase
diagram of the bimodal RFIM, in agreement with the early
predictions of mean-field theory~\cite{aharony78}. These authors
have also observed, at the disorder strength value $h=2.1$, using
standard and histogram Monte Carlo methods~\cite{hernandez97} and
more recently the WL algorithm~\cite{hernandez07}, the same
first-order-like characteristic dp structure and concluded that
there is a TCP at some intermediate value of the disorder
strength.

The existence of a dp structure in the canonical PDF is related to
a convex dip in the microcanonical entropy and it is known that
for some systems a mere observation of this structure is not
sufficient for the identification of a first-order transition. The
Baxter-Wu~\cite{baxter73,martinos05b,schreider05} and four-state
Potts models in 2D~\cite{wu82} are well-known examples of such
systems undergoing, in the thermodynamic limit, second-order phase
transitions. Recently, Behringer and Pleimling~\cite{behringer06}
have demonstrated for these two models that, the appearance of a
convex dip in the microcanonical entropy can be traced back to a
finite-size effect different from what is expected in a genuine
first-order transition. In other words, the pseudosignatures of a
first-order transition are finite-size effects, which can be
understood within a scaling theory of continuous phase transitions
and such first-order-like features cease to exist in the
thermodynamic limit. Similar first-order-like properties have been
observed in many other finite systems, such as the well-known
examples of the fixed-magnetization versions of the Ising model,
where it has been also shown that these finite-size effects
disappear in the thermodynamic
limit~\cite{pleimling01,binder03,martinos06}.

The present paper, is the first extensive numerical investigation
of this fundamental issue for the 3D bimodal RFIM. We proceed,
having in mind that a mere observation of a first-order structure
is not sufficient for the identification of the transition. This
is especially true for the present model, since its critical
behavior is obscured by strong and complex finite-size effects,
involving also the important issue of the lack of
self-averaging~\cite{wu06,dayan93,aharony96,wiseman98,parisi02,malakis06a}.
Thus, for a clear identification of the order of the transition,
we implement an appropriate version of the Lee-Kosterlitz (LK)
free-energy barrier method~\cite{lee90}. Initially, we used a
straightforward one-range (one-R) WL sampling on a set of a small
number of RF realizations, at two values of the disorder strength,
$h=2$ and $h=2.25$. This attempt enabled us to observe the
behavior of the free-energy barrier and the latent heat and
indicated that the transition remains continuous at the strong
disorder regime. By a second substantial attempt, using a combined
more efficient numerical scheme, we simulated large numbers of RF
realizations and verified that, indeed, the first-order-like
transition signatures are finite-size effects that disappear in
the thermodynamic limit.

The remainder of the paper is as follows: subsection~\ref{sec:2a}
gives a summary of the WL and Lee methods. In particular, we
explain, discuss, and give details of two different numerical
strategies, called hereafter as one-R approach and high-level
one-R approach. We continue, in subsection~\ref{sec:2b}, to
present the application of the LK free-energy barrier
method~\cite{lee90} on the numerical data, obtained via a
straightforward application of the one-R WL implementation on a
small ensemble of RF realizations for two values of the disorder
strength, $h=2$ and $h=2.25$. The same method is applied in
subsection~\ref{sec:2c} on the numerical data obtained via a new
proposal capable to simulate large numbers of RF realizations, at
the disorder strength value $h=2$. As will be explained in
subsection~\ref{sec:2a}, this latter strategy is an efficient and
accurate WL approach, which combines in three stages, the
multi-range (multi-R) WL algorithm, the high-level one-R WL
approach, and a final quite long Lee run, to obtain an
alternative, and presumably most accurate, density of states
(DOS). Subtle points behind the necessity of implementing such an
elaborate scheme will be discussed appropriately in the sequel.
Finally, we summarize our conclusions in section~\ref{sec:3}.

\section{Numerical schemes. Identification of the order of the transition}
\label{sec:2}

\subsection{Sampling the RFIM by the WL and Lee methods}
\label{sec:2a}

Several sophisticated simulation techniques, such as cluster
algorithms and flat-histogram approaches, have been used to study
the
RFIM~\cite{young85,machta00,hernandez97,hernandez07,dotsenko91,rieger93,rieger95,newman96},
while graph theoretical algorithms have been used to study
properties of the ground-states of this
model~\cite{ogielski86,swift97,auriac97,hartmann99,sourlas99,hartmann01,middleton02,wu06,dukovski03}.
Entropic sampling methods such as the Lee~\cite{lee93,lee06} and
WL~\cite{wang01a,wang01b,schulz03} methods are efficient
alternatives for complex systems and systems that undergo
first-order transitions. Accordingly, we will implement a
combination of such numerical approaches, based mainly on the WL
method, to study the nature of phase transition of the 3D bimodal
RFIM at the strong disorder regime. The WL algorithm is one of the
most refreshing improvements in Monte Carlo simulation and has
been applied to a broad spectrum of interesting problems in
statistical mechanics and
biophysics~\cite{lee06,yamaguchi01,troyer03,shell02,dayal04,adler04,zhou05,poulain06,zhou06,belardinelli07,malakis04,malakis07}.

To apply the WL algorithm, an appropriate energy range  of
interest has to be identified. A WL random walk (single spin flip)
is performed in this energy subspace. Trials from a spin state
with energy $E_{i}$ to a spin state with energy $E_{f}$ are
accepted according to the transition probability
\begin{equation}
\label{eq:3}p(E_{i}\rightarrow
E_{f})=\min\left[\frac{G(E_{i})}{G(E_{f})},1\right].
\end{equation}
During the WL process the DOS $G(E)$ is modified ($G(E)\rightarrow
f*G(E)$) after each spin flip trial by a modification factor
$f>1$. The WL iterative process ($j=1,2,\ldots$) is defined as a
process in which successive refinements of the DOS are achieved by
monotonically decreasing the modification factor $f_{j}$. Most
implementations use an initial modification factor
$f_{j=1}=e\approx 2.71828\ldots$ and a rule
$f_{j+1}=\sqrt{f_{j}}$, while a $5\%-10\%$ flatness criterion (on
the energy histogram) is applied in order to move to the next
refinement level ($j\rightarrow j+1$)~\cite{wang01a,wang01b}. The
WL process is terminated in a sufficiently high-level, at which
$f\approx 1$ (typically $f=1.000 \; 00001$). Note that the
detailed balanced condition is satisfied in the limit
$f\rightarrow 1$. There have been several papers in recent years
dealing with improvements and sophisticated implementations of the
WL iterative
process~\cite{lee06,yamaguchi01,troyer03,shell02,dayal04,adler04,zhou05,poulain06,zhou06,belardinelli07,malakis04,malakis07}.
Some of these suggestions appeared in studies of efficiency and
convergence of the WL iterative
process~\cite{lee06,troyer03,dayal04,zhou05,belardinelli07}, while
others were proposed in applications of the WL scheme in
simulating several models of statistical
mechanics~\cite{yamaguchi01,shell02,adler04,poulain06,zhou06,malakis04,malakis07}.
In our recent study, of the first-order transition of the
triangular SAF model~\cite{malakis07}, we also used a final stage
of an unmodified ($f=1$) Lee entropic simulation~\cite{lee93} by
applying after, a relatively long run, a Lee correction to an
already good approximation obtained by the WL process. This final
Lee entropic stage will be also followed here and it is hoped that
through this practice we improve accuracy, but also obtain an idea
of the level of approximation, since starting with a very accurate
DOS and using a sufficiently long run, the Lee correction should
produce an almost identical DOS.

In our implementation of this Lee entropic stage, we start with a
very good approximation of the DOS [$G_{WL}(E)$], obtained by the
WL process after a large number $n$ of WL iterations (we choose
$n=20$) in which we follow the above described reduction of the
modification factor $f$ ($f_{j=1}=e\approx 2.71828\ldots$,
$f_{j+1}=\sqrt{f_{j}}$, $j=1,2,\ldots,n$). This good estimate of
the DOS is used to determine the transition probabilities
[equation~(\ref{eq:3})] for an unmodified ($f=1$) random walk in
energy space, as described by Lee~\cite{lee93}, in a process which
obeys now the detailed balance condition (equation (5) of
reference~\cite{lee93}) and produces an almost flat energy
histogram $H(E)$ in the long run. Note that, a completely flat
histogram (besides statistical fluctuations) will be produced in
case one is using the exact DOS, as pointed out by
Lee~\cite{lee93}. However, provided that the Monte Carlo time is
long compared to the ergodicity time, we obtain a better estimate
for the DOS, i.e. $G_{Lee}(E)$, by the following prescription:
\begin{equation}
\label{eq:4}\log G_{Lee}(E)=\log G_{WL}(E)+\log H(E).
\end{equation}
The implementation described above is in fact very similar to the
suggestion of reference~\cite{lee06} for a repeated application of
the above Lee correction scheme, after a first stage of the WL
process consisting of $n$ WL iterations. Then, the successive Lee
corrections obtained by repeated applications of our equation
equation~(\ref{eq:4}) (equation (15) of reference~\cite{lee06})
will improve the original DOS, as shown in reference~\cite{lee06}.
This repeated application was started in~\cite{lee06} at an early
WL iteration level ($n=14$) and was tested favorably compared to
the simple WL process.

The multi-R WL approach is the implementation of the method in
which one splits the energy range in many
subintervals~\cite{wang01a}. This is almost a necessity for very
large lattices and the subintervals used are slightly overlapping.
The DOS's of the separate pieces are joined at the end of the
process. This multi-R approach is, of course, a much faster
process compared with a straightforward one-R implementation and
in many cases has provided very accurate results for very large
systems~\cite{wang01a}. Although, several papers have pointed out
problems with the accuracy, efficiency and convergence of the WL
method~\cite{lee06,troyer03,dayal04,belardinelli07}, several
important related questions are still unanswered, or at least, not
well understood. Possible distortions (systematic errors) induced
on the DOS by using a multi-R WL approach have not been adequately
discussed in the literature. Of course, boundary effects of WL
sampling in restricted energy subspaces (multi-R processes) were
observed, analyzed and successfully resolved in
reference~\cite{schulz03}. However, subtle effects, coming from
the inevitable breaking of the ergodicity of phase space, may be
inherent in any restricted energy subspace or multi-R method.
Below, we will present a novel case coming from our recent studies
of the RFIM. An escape from this novel, rather discouraging case,
will be proposed by compromising between the multi-R and the one-R
approach.

\begin{figure}[ht]
\centerline{\includegraphics*[width=12 cm]{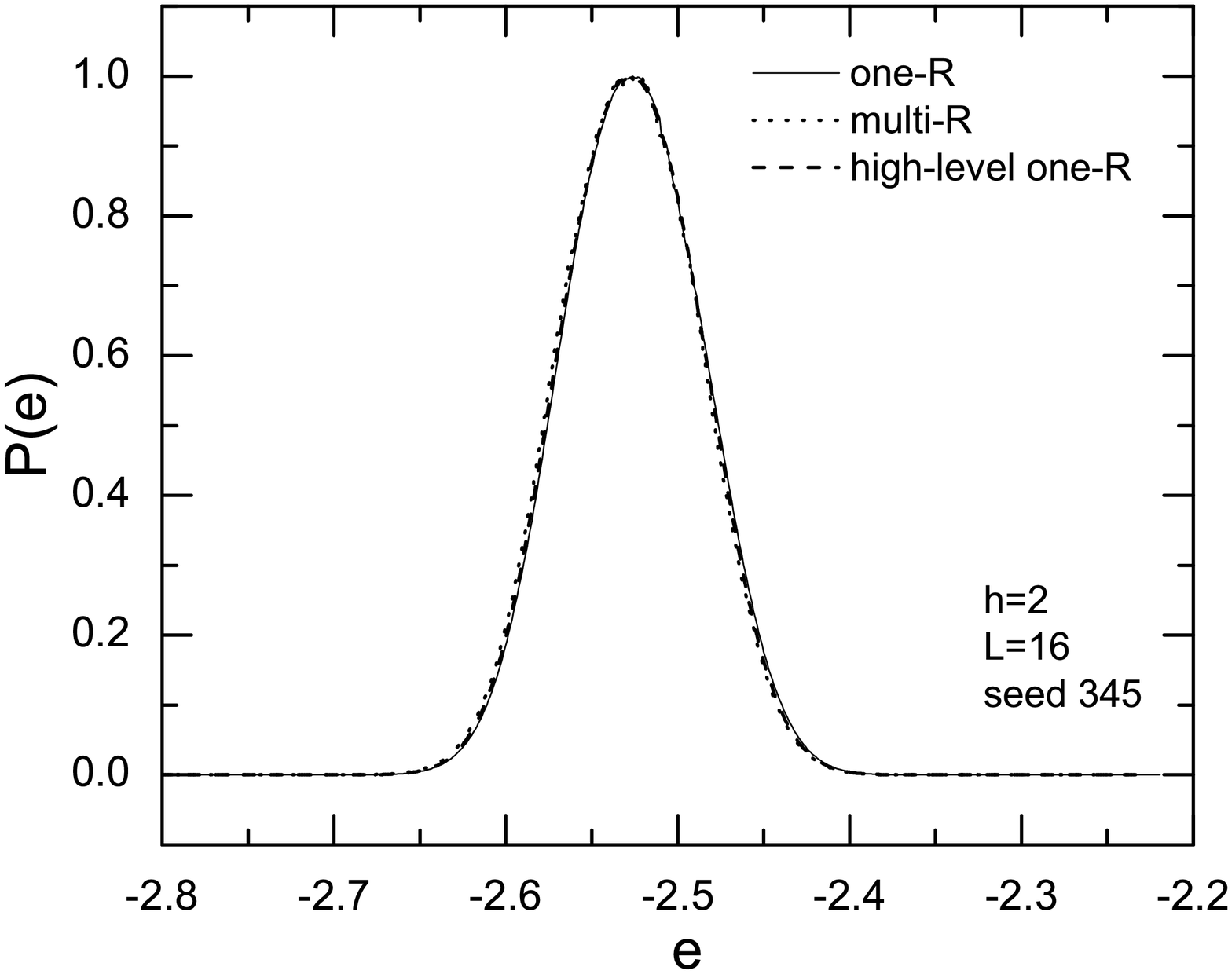}}
\caption{\label{fig:2}A typical sp energy PDF for a RF realization
at the temperature corresponding to the specific heat peak. Three
almost coinciding energy PDF's are shown, corresponding to the
three approaches discussed in the text, i.e. one-R, multi-R, and
high-level one-R. The PDF's are expressed as a function of the
energy per site $e(=E/N)$.}
\end{figure}
The WL method has been already applied to the RFIM in several
previous studies. Two such recent investigations, directly related
to this work, have been presented for the Gaussian~\cite{wu06} and
the bimodal RFIM~\cite{hernandez07}. As pointed out in the
introduction, both of these studies have observed and discussed
first-order-like properties of the RFIM at the strong disorder
regime. The WL method was also implemented, in restricted energy
subspaces, for the study of the bimodal RFIM in our earlier
studies~\cite{malakis06a}. In these papers, a systematic
restriction of the energy space, with increasing the lattice size,
was used and explained in detail in order to further improve the
efficiency of the WL method. This approach followed the general
spirit of our earlier proposal of estimating the critical behavior
of classical statistical systems via entropic simulation in
dominant energy subspaces. This restrictive version, utilizes the
so called critical minimum energy subspace (CrMES)
technique~\cite{malakis04} to locate and study finite-size
anomalies of systems by carrying out the random walk only in the
dominant energy subspaces. Generally, our finite-size scaling
studies have shown that this restrictive practice can be followed
in systems undergoing second-order~\cite{malakis04} and also
first-order transitions~\cite{malakis07}. Furthermore, in our
recent study of the phase diagram of the 3D bimodal
RFIM~\cite{fytas07} we have used a one-R and looser version of
this restrictive scheme. In this case we have used the high-levels
of the one-parametric WL method as a convenient entropic vehicle,
by which the accumulation of the two-parametric, exchange-energy,
field-energy, histograms would provide, via extrapolation, a good
approximation for the two-parameter DOS necessary to find several
points of the phase diagram.
\begin{figure}[ht]
\centerline{\includegraphics*[width=16 cm]{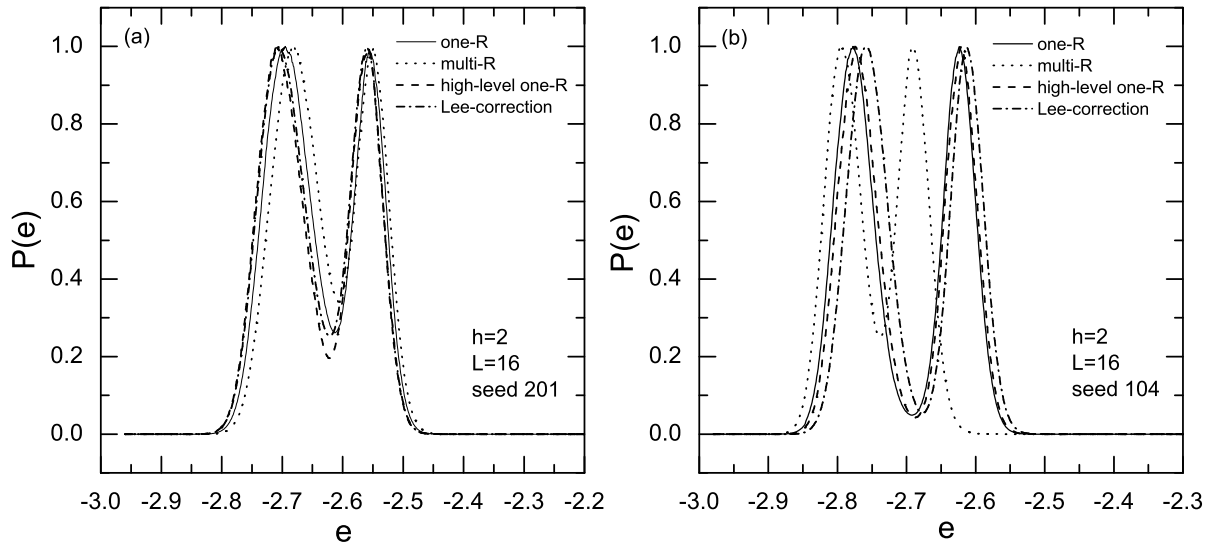}}
\caption{\label{fig:3}Different energy PDF's for two RF
realizations estimated by the numerical approaches discussed in
the text: one-R, multi-R, high-level one-R, and final Lee. The
PDF's are determined at the temperature where the two peaks are of
equal height. Note that, only the RF realization of panel (b)
shows a large distortion of the dp structure in the case of the
multi-R approach, when compared to the other (one-R) schemes.}
\end{figure}
Since substantial histogram accumulation is necessary to overcome
statistical errors in such an application, the faster multi-R
approach was not used and for having a reliable extrapolation
scheme the energy spectrum for the simulation was restricted only
from the high-energy side, while the entire low-energy part of the
spectrum down to the ground-state was included. For the
restriction of the high-energy side we used our data from our
previous study of the model at the value $h=2$. However, for the
larger lattice sizes, one can conveniently avoid the ground-state
neighborhood.

In the present study, we initially used the data of this last
straightforward one-R approach~\cite{fytas07} to observe the
behavior of the dp structure of some typical RF realizations at
the strong disorder regime ($h=2$ and $h=2.25$). As already
pointed out, these results appear in subsection~\ref{sec:2b} and
are obtained by using a final of $j=20$ WL iteration levels for
the smaller lattices up to $L=16$ and a final of $j=24$ WL
iteration levels for the larger sizes ($L\geq 20$).

Subsequently, and in order to simulate larger numbers of dp RF
realizations, we decided to test carefully and then use a multi-R
approach. Thus, we compared the (energy PDF) dp's of some typical
RF realizations obtained by the one-R approach with the dp's
obtained by a usual multi-R approach (corresponding to the same or
an even higher level of the WL process). For some RF realizations
the energy PDF graphs almost coincided and this was especially
true for the single-peak (sp) RF realizations. But also for
several dp RF realizations the corresponding graphs were close
enough and within statistical errors. Figures~\ref{fig:2} and
\ref{fig:3}(a) show two such examples, one corresponding to a sp
RF realization and one to a dp RF realization. However, for some
peculiar dp RF realizations the dp graphs resulting form the
multi-R approach were dislocated and with a rather large deviation
in their depth, when compared with the dp energy PDF graphs
obtained for the same RF realizations by the one-R approach.
Figure~\ref{fig:3}(b) shows a characteristic case, corresponding
to a serious dislocation and underestimation of the dp structure.
Furthermore, it was observed that the dp details were very
sensitive to the division of the energy range to subintervals,
indicating that the distortion errors were due to the application
of the multi-R approach. After several tests, we concluded that
this peculiar problem is related to the division of the dp range
in subintervals. It appears that for some RF realizations, the
structure of the convex dip in the microcanonical entropy is not
well estimated by using the multi-R approach within the dp range.
We concluded that the details of the convex dip are sensitive to
possible subtle violations of ergodicity, induced by the multi-R
approach. Thus, we tried to find an alternative that will not
suffer from this problem and still be efficient enough so that we
could simulate large numbers of RF realizations. The developed
method will be called high-level one-R WL approach and is a
further sophistication in the same spirit of our earlier practice
in optimizing the WL entropic sampling. It combines the multi-R
and the one-R approaches in an almost optimum way and seems to
meet the needs of a careful estimation of the dp structure of the
present model.

The details of this approach, applied here only for $h=2$, are as
follows. For each lattice size, a wide energy subspace restricted
mainly from the high-energy side is divided in relatively small
subintervals, of the order of $100$ energy levels and a multi-R
approach is applied up to the $j=16$ WL iteration level. This
completes the first stage of the approach and the DOS obtained is
used to estimate for each particular RF realization the dp range.
This identification is easily achieved by using our earlier
practice for first-order transitions~\cite{malakis07}, by finding
the appropriate temperature of equal height for the two peaks of
the energy PDF. The energy PDF at this temperature is normalized
so that the height of the two peaks is unity and corresponds to
energies $E_{1}$ and $E_{2}$. The dp range is now identified as
follows: the left-end of the dp subspace is the energy $E_{1-}$,
for which the density becomes greater than $10^{-6}$ starting from
$E_{min}$ and respectively the right-end of the dp subspace is the
energy $E_{2+}$, for which also the density becomes greater than
$10^{-6}$ starting from $E_{max}$. Having this first approximate
identification of the dp subspace, a one-R WL walk is again
performed at the level $j=16$ in a subspace which is wider than
the dp subspace by a factor of $10\%$ at each end. In other words,
the left-end $E_{1-}$ is shifted to the left by $10\%$ of the dp
range and correspondingly the right-end $E_{2+}$ is shifted to the
right by the same amount. After the $j=16$ one-R approach the ends
of the dp subspace are re-estimated and fixed. The one-R WL
approach is then carried on only in this dp subspace for the
higher levels $j=17,18,19$, and $j=20$. This completes the second
stage of our approach. Finally, an unmodified Lee random walk is
performed in this dp subspace, using the last approximation of the
WL DOS for the transition rates. The Lee correction is applied at
the end to produce an alternative estimate for the DOS. The time
duration of this last Lee run is taken to be equal to the duration
of the four last one-R WL iterations. For some RF realizations,
this one-R process was pushed up to $j=24$ to observe differences
and estimate statistical errors. In all cases, these statistical
errors were very small, much smaller than the observed
sample-to-sample fluctuations (see also discussion in
subsection~\ref{sec:2c} below).

We give here some details for the sizes of the dp subspaces
involved in the above scheme. For $L=24$, the initial restricted
energy subspace used for the multi-R process was of the order of
$2600$ energy levels, i.e. counting energy levels from the all
minus spin state this was the subspace defined by the levels
$ie=100$ to $ie=2700$. Typically the size of the resulting dp
subspace was of the order of $800$ energy levels, which is about
$30\%$ of the initial restricted energy subspace. The left-end
$E_{1-}$ roughly fluctuated, for a sample of $100$ RF
realizations, between the levels $ie=600-900$, while the right-end
$E_{2+}$ between the levels $ie=1400-1700$. Respectively, for
$L=32$ the initial restricted energy subspace used for the multi-R
process was of the order of $4200$ energy levels, defined by the
levels $ie=300$ to $ie=4500$. Typically, the size of the resulting
dp subspace was again of the order of $800$ energy levels, which
is about $20\%$ of the initial restricted energy subspace. In this
case, $E_{1-}$ fluctuated, again for a sample of $100$ RF
realizations, between the levels $ie=1450-1950$, while $E_{2+}$
between the levels $ie=2250-2750$.

To conclude the above remarks, let us point out that, for a
typical RF realization, when $L=24$, a safe dp location is
established after the $j=16$ one-R level and consists of about
$960$ energy levels ($800+20\%\times 800$). It is quite
astonishing that the same energy space requirements are needed
also for $L=32$ and this is related to the final conclusion of
this paper, that the dp peak width, in units of energy per site,
tends to zero in the limit $L\rightarrow \infty$. The above
remarks clarify also the reasons behind the efficiency of the
present proposal (high-level one-R approach). Typically, for one
RF realization of a lattice size $L=16$ at the disorder strength
value $h=2$ (figure~\ref{fig:3}), the simulation time $t$ for the
one-R WL process, in all the energy subspace ($E_{min}, E_{max}$),
was of the order of $12$ hours performed in a Pentium IV 3GHz. The
simulation times corresponding to the other cases presented in
panel (b) of figure~\ref{fig:3} are as follows: multi-R approach
$t/24$ and high-level one-R WL approach together with the final
Lee run $t/9$. We may also point out, that very recently,
Fern\'{a}ndez \etal~\cite{fernandez08}, have found that the phase
space for the first-order transition of the 3D site-diluted
four-states Potts model is reduced, as compared with the
expectations from simulations in small lattice sizes, a behavior
very similar to the above observations. Their microcanonical
approach~\cite{fernandez08,mayor07} may also be an interesting
alternative, not used previously, for the study of the present
model.

\subsection{One-R WL approach. Transition-identification by the LK method}
\label{sec:2b}

In this subsection, we present the application of the LK method on
the numerical data obtained by the one-R WL method. As mentioned
in the previous subsection, the one-R approach on both values of
the disorder strength ($h=2$ and $h=2.25$) was applied in a wide
energy spectrum and we have conveniently avoided a suitable
ground-state neighborhood. The total number of RF realizations
simulated ($N_{tot}$) varies from $20$ realizations for $L\leq 24$
to $10$ realizations for $L>24$.

From a traditional point of view, the nature of a phase transition
can be, in principle, determined by examining the finite-size
scaling of various thermodynamic quantities, such as the specific
heat and susceptibility peaks. These two quantities, as well as
others, are expected, from the general theory of first-order
transition, to follow an $L^{d}$ divergence and may be used as
indicators of the order of the transition. However, this practice
is very often inconclusive even for simple systems and may be
seriously questioned for random systems in which fundamental and
subtle problems exist concerning the averaging process over
disorder. For the present RFIM, the lack of self-averaging
observed by the present authors~\cite{malakis06a} (see also
references~\cite{wu06,dayan93,parisi02}), may be of crucial
importance especially in trying to construct convenient indicators
for the nature of the transition at the strong disorder regime.
First-order-like realizations are expected to exhibit sharp
specific heat and susceptibility peaks, such as that observed also
in the Gaussian case studied by Wu and Machta~\cite{wu06}.
Therefore, according to our previous papers~\cite{malakis06a} and
as pointed out also in reference~\cite{hernandez07} the
information concerning an individual first-order-like realization
will be washed out - as a result of the strong fluctuation in the
pseudocritical temperature - in considering for instance the
behavior of the average specific heat curve.

From the above discussion, it is obvious that in order to avoid
problems with the lack of self-averaging property, the
first-order-like features of each realization must be computed
separately and the disorder average should be applied at the end
in the proper first-order indicator. The most convenient approach
in this case is to use the free-energy barrier method proposed by
Lee and Kosterlitz~\cite{lee90}.
\begin{figure}[ht]
\centerline{\includegraphics*[width=16 cm]{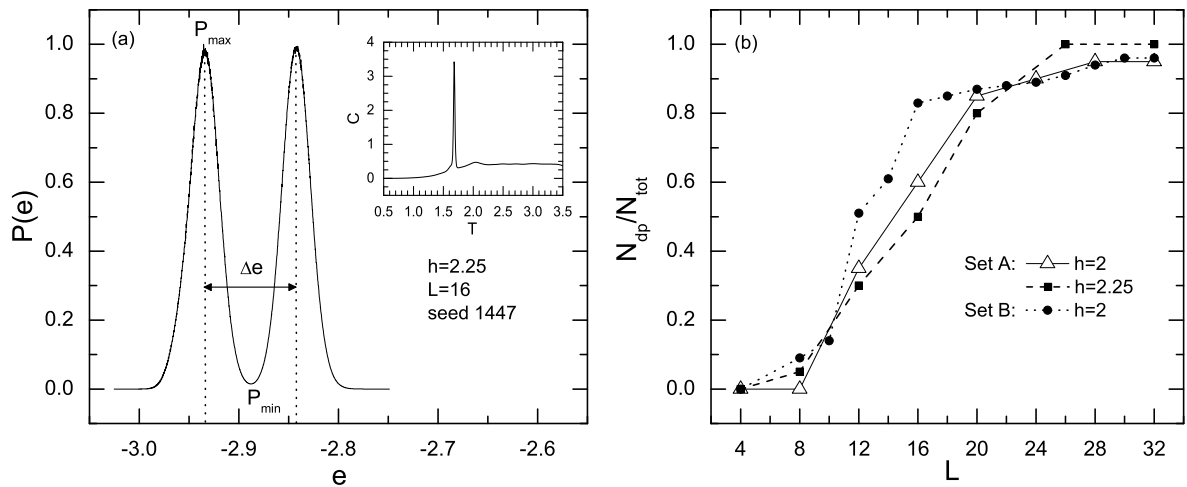}}
\caption{\label{fig:4}(a) Energy PDF at the temperature where the
two peaks are of equal height for a first-order-like RF
realization of a lattice size $L=16$ at disorder strength
$h=2.25$. The inset shows the corresponding sharp specific heat
peak. (b) Ratio $N_{dp}/N_{tot}$ of realizations showing a dp
energy PDF in an ensemble of $N_{tot}$ number of realizations as a
function of the linear system size $L$. Set A: $\{h=2;\;
L=4,8,12,16,20,24,28,32 \}$ (open triangles) and $\{h=2.25;\;
L=4,8,12,16,20,26,32 \}$ (filled squares). Set B: $\{h=2;\;
L=4,8,10,12,14,16,18,20,22,24,26,28,30,32 \}$ (filled circles).}
\end{figure}
This method has been already successfully applied by Chen
\etal~\cite{chen92} for the study of an analogous disordered
system, namely the two-dimensional eight-state Potts model with
quenched bond randomness. Thus, we will proceed now to apply the
LK method for the identification of the transition of the bimodal
RFIM at the strong disorder regime. The method is well-known and
has been widely applied to several spin
models~\cite{lee90,malakis07,chen92}, so we will proceed giving
only the necessary definitions adapted to the present disordered
system.

Figure~\ref{fig:4}(a) illustrates, in the main frame, the typical
dp energy probability distribution (seed $1447$) and in the inset
the corresponding sharp specific heat peak of a first-order-like
realization on a cubic lattice of linear lattice size $L=16$ at
the disorder value $h=2.25$. With the help of this figure, let us
define the surface tension $\Sigma(L)=\Delta F(L)/L^{d-1}(=\Delta
F(L)/L^{2})$ for each dp realization, where the definition of the
LK free-energy barrier is, using the canonical energy PDF $P(e)$,
$\Delta F=k_{B}T\ln{[P_{max}/P_{min}]}$ ($e=E/N$, as in
figure~\ref{fig:2}). Therefore, with the help of only the
generated dp realizations ($N_{dp}$ is their number, see also the
discussion below), we define the disorder average of the surface
tension, which is proposed as the relevant indicator representing
the ensemble of dp realizations, as
\begin{equation}
\label{eq:5}
[\Sigma(L)]_{av}=\frac{1}{N_{dp}}\sum_{i=1}^{N_{dp}}\Sigma_{i}(L).
\end{equation}
The second important ingredient in the size development of the
observed first-order-like properties of the 3D bimodal RFIM is the
behavior of the width $\Delta e(L)$ of the individual dp's,
representing the latent heat of the transition, in the case of a
first-order transition. Again with the help of the illustration in
figure~\ref{fig:4}(a), we define the disorder average over the
ensemble of dp realizations of the width of the transition as
\begin{equation}
\label{eq:6} [\Delta
e(L)]_{av}=\frac{1}{N_{dp}}\sum_{i=1}^{N_{dp}}\Delta e_{i}(L).
\end{equation}

Figure~\ref{fig:4}(b) presents the relative number of such
first-order-like realizations $N_{dp}$, in a total number of
$N_{tot}$ realizations. Set A refers to the straightforward one-R
WL approach for the two values $h=2$ and $h=2.25$ of the disorder
strength. The observed increase with lattice size of the
probability for such first-order-like realizations, is in
qualitative agreement with the general behavior reported by Wu and
Machta~\cite{wu06} for the Gaussian RFIM. From Table V of
reference~\cite{wu06} one observes a general tendency of the ratio
$N_{dp}/N_{tot}$ to increase with the system size and the disorder
strength and since we are studying the system at a relatively
higher disorder strength value, our ratios are quite comparable
with those given in Table V of reference~\cite{wu06}, although the
latter refer to the Gaussian RFIM. Clearly, at the strong disorder
regime and as the lattice size increases, the percentage of
realizations showing a dp in the energy probability distribution
increases and approaches unity very rapidly.
\begin{figure}[ht]
\centerline{\includegraphics*[width=16 cm]{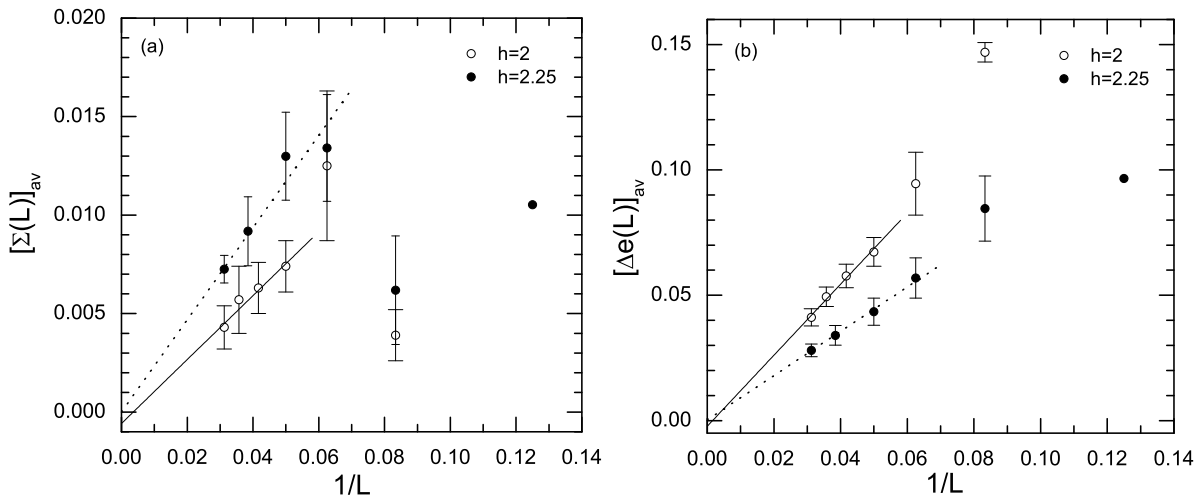}}
\caption{\label{fig:5}$1/L$-behavior of $[\Sigma(L)]_{av}$ (a) and
$[\Delta e(L)]_{av}$ (b) at $h=2$ (open circles) and $h=2.25$
(filled circles) for $L\geq 8$. Sample-to-sample fluctuations are
illustrated by the error bars. The solid and dotted lines are
corresponding linear fits for the larger sizes ($L\geq 16$). In
both panels and for both values of the disorder strength, a
limiting value very close to zero for the free-energy barrier and
the latent heat is obtained, indicating a continuous transition.}
\end{figure}
Thus, in our case for lattice sizes of the order of $L\geq 24$
almost all of the simulated realizations showed a dp energy
probability distribution (in fact for the value $h=2.25$ the ratio
$N_{dp}/N_{tot}$ reached unity). This observation explains why we
have used only the $N_{dp}$ realizations in the disorder averaging
(equations~(\ref{eq:5}) and (\ref{eq:6})), since it implies that
only these realizations are of interest, since for large lattices
these will dominate the behavior. This practice avoids transient
effects, coming from the small lattices, and we have further
pushed it by applying here a quite rather strict criterion for the
definition of the dp realizations: $P_{min}<0.75$ (note the
normalization of the energy PDF: $P_{max}=1$ in
figure~\ref{fig:4}(a)).

The behavior of the disorder average of the surface tension
$[\Sigma(L)]_{av}$, for both $h=2$ and $h=2.25$, as a function of
the inverse lattice size, is shown in figure~\ref{fig:5}(a).
Linear fits are applied only for the data corresponding to sizes
$L\geq 16$. From these linear plots (solid and dotted lines) it
appears that $[\Sigma(L)]_{av}$ approaches zero, as expected at a
second-order transition. This observation strongly indicates that,
what we are observing from these dp realizations for small sizes
is a finite-size effect that will disappear in the thermodynamic
limit. The solid and dotted lines explicitly illustrate this,
using a linear extrapolation of the large size data for $h=2$ and
$2.25$, giving an almost zero surface tension in the limit
$L\rightarrow \infty$ for both values of the disorder strength:
$0.0006\pm 0.009$ and $0.000009\pm 0.03$, respectively.
Furthermore, figure~\ref{fig:5}(b) depicts an undeniable steady
approach to zero of the above representative width, again for both
values of the disorder strength $h=2$ and $h=2.25$. The linear
extrapolation attempts are shown by the solid and dotted lines and
give also an almost zero value for the latent heat of the order of
$-0.002\pm 0.004$ and $0.0002\pm 0.008$, respectively. This is a
further strong manifestation in favor of the continuous phase
transition scenario. Thus, the evidence presented in this
subsection for the 3D bimodal RFIM are in agreement with the
favored view of most existing theoretical and numerical
studies~\cite{gofman93,falicov95,malakis06a,rieger93} that the
phase transition of the 3D RFIM is of second-order. In order to
present even stronger numerical evidence of a vanishing (in the
limit $L\rightarrow \infty$) surface tension we will now attempt
to go well beyond the observation of several typical RF
realizations. In the next subsection, numerical evidence will be
presented for the finite-size behavior of the free-energy barrier
and the latent heat of much larger ensembles of RF realizations,
obtained via the efficient high-level one-R entropic scheme
described in subsection~\ref{sec:2a}.

\subsection{Revisiting the order of the transition by the high-level one-R WL approach}
\label{sec:2c}

In this subsection, we will apply the LK method on the numerical
data obtained by our second numerical strategy, described in
subsection~\ref{sec:2a}. Using the high-level one-R WL approach
and its final Lee correction, we generated numerical data for
large number of RF realizations at the disorder strength value
$h=2$. In this case, the number $N_{tot}$ of realizations varied
so that for every lattice size $L>10$, $100$ RF realizations
showing a dp structure in the energy PDF ($N_{dp}=100$) were
simulated. For the small sizes $L=8$ and $L=10$ only $10$ dp
realizations have been identified in respective ensembles of
$N_{tot}=112$ and $N_{tot}=74$ simulated realizations.

Set B in figure~\ref{fig:4}(b) refers to the above mentioned large
ensembles of realizations simulated at $h=2$. In the present case,
a looser criterion ($P_{min}<0.9$) was applied for the
identification of a dp realization. The corresponding ratios
$N_{dp}/N_{tot}$ in figure~\ref{fig:4}(b) almost coincide for set
A and set B, for the larger sizes. This is easily explained by
observing that for large sizes almost all dp realizations appear
to have a quite deep minimum in the energy PDF. However, as it
will be shown below, the scaling of these minima will not support
a first-order character of the transition.
\begin{figure}[ht]
\centerline{\includegraphics*[width=16 cm]{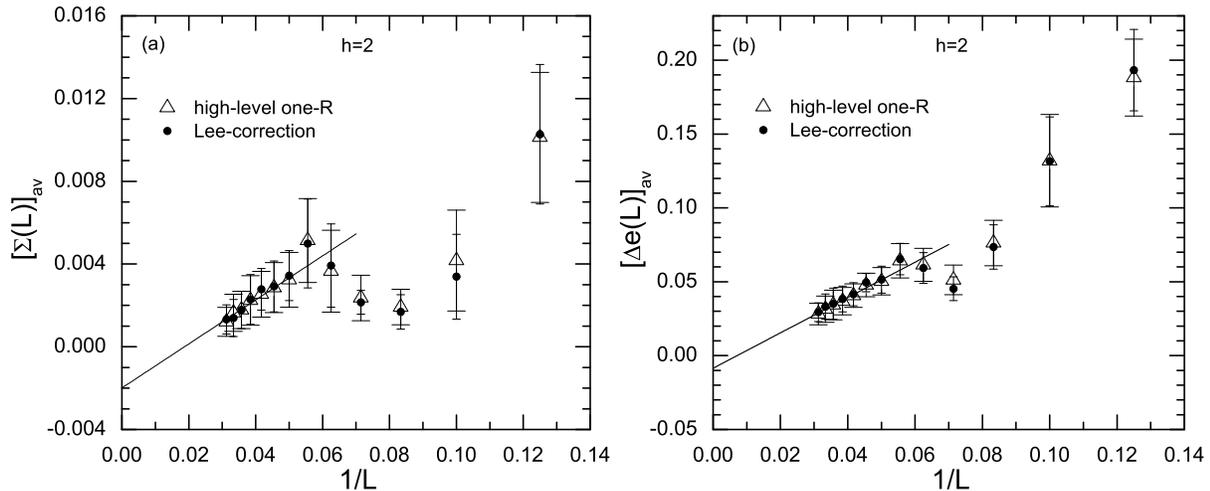}}
\caption{\label{fig:6}$1/L$-behavior of $[\Sigma(L)]_{av}$ (a) and
$[\Delta e(L)]_{av}$ (b) at $h=2$ and for $L\geq 8$. Two set of
results are shown, corresponding to the high-level one-R WL
approach (open triangles) and the Lee correction (filled circles).
The sample-to-sample fluctuations for the case of the high-level
one-R WL results are shown with the larger cap width. The solid
lines are the corresponding linear fits for $L\geq 16$, giving
very small, close to zero, negative values for
$[\Sigma(L\rightarrow \infty)]_{av}$ and $[\Delta e(L\rightarrow
\infty)]_{av}$, thus verifying the scenario of
figure~\ref{fig:5}.}
\end{figure}

Figure~\ref{fig:6} presents our results for the disorder averaged
surface tension $[\Sigma (L)]_{av}$ and latent heat $[\Delta
e(L)]_{av}$ over set B of realizations at $h=2$. The open
triangles refer to the results obtained by the high-level one-R WL
approach, whereas the filled circles to those estimated from the
final Lee correction. In panel (a) the values of $[\Sigma
(L)]_{av}$ are shown for $L\geq 8$. Although for sizes up to
$L=18$ $[\Sigma(L)]_{av}$ seems to steadily increase, for sizes
$L\geq 20$ a clear approach to zero is observed and this fact is
compatible to the behavior of figure~\ref{fig:5}(a). Respectively,
panel (b) shows the values of $[\Delta e(L)]_{av}$, also for
$L\geq 8$. The final large size decrease, is explicitly
illustrated by the solid line (in both panels of
figure~\ref{fig:6}), revealing the true asymptotic behavior of
these quantities. Specifically, the solid lines are linear fits
performed on the data obtained by the high-level one-R WL
approach, for the larger lattice sizes studied ($L\geq 16$),
giving: $-0.00171\pm 0.0009$ and $-0.00852\pm 0.005$ for $[\Sigma
(L\rightarrow \infty)]_{av}$ and $[\Delta e(L\rightarrow
\infty)]_{av}$, respectively. As it can be seen from this figure,
the data obtained by the Lee correction process would practically
provide the same limiting values for the surface tension and the
latent heat and therefore the corresponding linear fits are not
shown. Noteworthy that, a comparison between statistical errors
and sample-to-sample fluctuations is quite apparent in this
figure. The values of $[\Sigma(L)]_{av}$ and $[\Delta e(L)]_{av}$
estimated over the two sets of realizations (set A and set B) for
$h=2$ estimated via the two different numerical strategies
presented in this paper are of the same order. Those of set A are
slightly larger as a consequence of the more strict criterion,
$P_{min}<0.75$, used for the identification of the dp RF
realizations.

The interesting first-order-like properties of the model, reported
by Hern\'{a}ndez and Diep~\cite{hernandez97} and Hern\'{a}ndez and
Ceva~\cite{hernandez07} for the bimodal RFIM and by Wu and
Machta~\cite{wu06} for the Gaussian RFIM, have added more
complication and novelty to the RFIM. The first-order-like
characteristics of the Gaussian RFIM found by Wu and
Machta~\cite{wu06} revealed that the appearance of these strong
finite-size effects are independent of the RF distribution and
their existence is related to the value of the disorder strength.
This observation is not compatible with mean-field theory, since
its first-order prediction for only the bimodal case depends on
the existence of a minimum at zero-field of the
distribution~\cite{aharony78}. The present study has illustrated
that these characteristics are most likely effects complicating
the finite-size behavior of the model but not determining its true
asymptotic scaling behavior.

Our results clearly indicate that the interface tension $[\Sigma
(L)]_{av}$ vanishes and the two peaks of the energy PDF move
together in the thermodynamic limit and therefore provide
convincing evidence that the transition is continuous and that
there in no TCP along the phase transition line. Consequently, the
coexistence between an ordered phase and a disordered phase will
be hardly detectable in the thermodynamic limit. Nevertheless, for
large but finite systems, the dip represents a considerable
barrier between the ordered phase and the disordered one, so that
in some sense, one may speak for a phase coexistence for finite
systems. The results of figures~\ref{fig:5}(a) and \ref{fig:6}(a)
indicate a linear approach of the interface tension to zero in the
limit $L\rightarrow\infty$, and therefore an exponential increase
of the ratio $P_{max}/P_{min}$ in $L$, and point to an
unconventional continuous transition, in which the energy PDF will
approach two delta functions that move together (see
figures~\ref{fig:5}(b) and \ref{fig:6}(b)) in the thermodynamic
limit. Such an unconventional behavior has been first predicted by
Eichhorn and Binder~\cite{eichhorn96}, for the the order-parameter
PDF of the 3D random-field three-state Potts model. These authors
have explained such an unusual behavior by presenting in detail
the consequences of a scenario (including leading corrections to
scaling) based on a finite-size scaling statement for the
order-parameter universal PDF. According to this scenario, the
finite-size scaling behavior in RF systems can be recovered and
the relative width of the corresponding order-parameter PDF peaks
vanishes in the scaling limit as $L^{-\theta/2}$, where $\theta$
is the critical exponent describing the violation of hyperscaling
($2-\alpha=(d-\nu)\theta$). In conclusion, the results of this
paper and the observations of
references~\cite{wu06,hernandez97,hernandez07} are strong
indications of a similar unusual scenario for a continuous
transition, calling for further investigation, such as the
determination of the exponent $\theta$.

\section{Conclusions}
\label{sec:3}

Two entropic sampling numerical strategies have been implemented
for the study of the first-order-like properties of the 3D bimodal
RFIM. Our experience and comparative studies, using different
numerical approaches, revealed the sensitivity of the double-peak
structure of the energy probability density function of the model,
especially with increasing the system size. Thus, the need for
careful implementations of entropic sampling techniques in cases
of complex systems has been critically discussed. An efficient
high-level one-range Wang-Landau approach has been proposed as a
quite safe alternative, avoiding subtle problems related to the
position and the depth of the minima of the double-peak energy
probability density functions.

Reliable data were obtained using this high-level one-range
Wang-Landau approach and the corresponding Lee correction for
large numbers of random-field realizations and quite large lattice
sizes, up to $L=32$. Using these data and by a systematic
finite-size analysis, implementing the Lee-Kosterlitz method, we
have studied the nature of the transition at the strong disorder
regime. Our results for both the free-energy barrier and latent
heat for the ensemble of double-peak random-field realizations
suggest a behavior in accordance with a continuous transition.
These results disclose the open controversy for the existence of a
tricritical point in the phase diagram of the 3D bimodal RFIM at
the strong disorder regime and serve in favor of the unusual
scenario for a continuous transition, originally proposed by
Eichhorn and Binder~\cite{eichhorn96}. It will be interesting to
repeat the present investigation for the wide bimodal distribution
(with a Gaussian width) and even for the Gaussian distribution, at
the strong disorder regime, since this would provide additional
confidence to our conclusions.

\ack{The authors would like to thank Professor A N Berker for
useful discussions. Financial support by EPEAEK/PYTHAGORAS under
Grant No. $70/3/7357$ is gratefully acknowledged. N G Fytas was
supported by the Alexander S. Onassis Public Benefit Foundation.}

\end{document}